\documentclass[twocolumn,showpacs,floatfix,superscriptaddress,amsmath,amssymb,prl]{revtex4}
\usepackage{mathrsfs}
\usepackage{txfonts}
\usepackage{amssymb}
\usepackage{graphicx}
\usepackage{hyperref}
\usepackage{ulem}
 \usepackage{overpic}
 \usepackage{psfrag}
 \newcommand{\be}{\begin{equation}}
\newcommand{\ee}{\end{equation}}

\begin{document}
\title{Spontaneous symmetry breaking and bifurcations in ground state fidelity for quantum lattice systems}

\author{Jian-Hui Zhao}
\affiliation{Centre for Modern Physics and Department of Physics,
Chongqing University, Chongqing 400044, The People's Republic of
China}
\author{Hong-Lei Wang}
\affiliation{Centre for Modern Physics and Department of Physics,
Chongqing University, Chongqing 400044, The People's Republic of
China}
\author{Bo Li}
\affiliation{Centre for Modern Physics and Department of Physics,
Chongqing University, Chongqing 400044, The People's Republic of
China}
\author{Huan-Qiang Zhou}
\affiliation{Centre for Modern Physics and Department of Physics,
Chongqing University, Chongqing 400044, The People's Republic of
China}

\begin{abstract}
Spontaneous symmetry breaking occurs in a system when its
Hamiltonian possesses a certain symmetry, whereas the ground state
wave functions do not preserve it.  This provides such a scenario
that a bifurcation, which breaks the symmetry, occurs when some
control parameter crosses its critical value. It is unveiled that
the ground state fidelity per lattice site exhibits such a
bifurcation for quantum lattice systems undergoing quantum phase
transitions. The significance of this result lies in the fact that
the ground state fidelity per lattice site is \textit{universal}, in
the sense that it is model-independent, in contrast to
(model-dependent) order parameters. This fundamental quantity may be
computed by exploiting the newly-developed tensor network algorithms
on infinite-size lattices. We illustrate the scheme in terms of the
quantum Ising model in a transverse magnetic field and the spin
$1/2$ XYX model in an external magnetic field on an infinite-size
lattice in one spatial dimension.

\end{abstract}


\maketitle

 {\it Introduction.} Quantum phase transitions (QPTs)~\cite{sachdev,wen} arise from the
cooperative behaviors in quantum many-body systems,  in which
long-range orders emerge. In the conventional Landau-Ginzburg-Wilson
paradigm, the most fundamental notion is spontaneous symmetry
breaking (SSB), with the symmetry-broken phase characterized by the
nonzero values of a local order parameter. An SSB occurs in a system
when its Hamiltonian enjoys a certain symmetry,  whereas the ground
state wave functions do not preserve it~\cite{anderson,coleman}. The
implication of an SSB is two-fold: first, a system has stable and
degenerate ground states, each of which breaks the symmetry of the
system; second, the symmetry breakdown results from random
perturbations. This leads to such a scenario that a bifurcation,
which breaks the symmetry, occurs, when some control parameter
crosses its critical value. Conventionally, this is reflected in
local order parameters.

The latest advances in our understanding of QPTs originate from the
perspectives of both entanglement~\cite{book} and
fidelity~\cite{zp,zhou,zov,fidelity}, which are basic notions in
quantum information science. In Refs.~\cite{zhou,zov}, it has been
argued that the ground state fidelity per lattice site is
\textit{fundamental} in the sense that it may be used to
characterize QPTs, regardless of what type of internal order is
present in quantum many-body states. The argument is solely based on
the basic Postulate of Quantum Mechanics on quantum measurements,
which implies that two non-orthogonal quantum states are not
reliably distinguishable~\cite{nielsen}. In other words, the ground
state fidelity per lattice site is able to describe QPTs arising
from an SSB and/or topological order~\cite{statement}. This has been
further confirmed in Refs.~\cite{zhou1,zhou2}, where topologically
ordered states in the Kitaev model on the honeycomb lattice and the
Kosterlitz-Thouless phase transition are investigated from the
fidelity perspective, respectively. Moreover, even for systems with
symmetry-breaking orders, it is advantageous to adopt the ground
state fidelity per lattice site instead of using the conventional
local order parameters, due to the fact that it is
model-independent, although one may systematically derive local
order parameters from tensor network (TN) representations of quantum
many-body ground state wave functions by investigating the reduced
density matrices for local areas on an infinite-size
lattice~\cite{zhou3}. However, it remains unclear whether or not it
is possible for the ground state fidelity per lattice site to
capture bifurcations arising from an SSB.

In this Letter, we attempt to fill in this gap. First, we
demonstrate that the newly-developed TN algorithms on infinite-size
lattices may produce degenerate ground states arising from an SSB,
each of which results from a randomly chosen initial state subject
to an imaginary time evolution. Second, it is unveiled that an SSB
is reflected as a bifurcation in  the ground state fidelity per
lattice site for quantum lattice systems undergoing QPTs with
symmetry-breaking orders~\cite{bifurcation}.  The significance of
this conclusion lies in the fact that, on the one hand, this
establishes the connection between the fidelity approach to QPTs and
the singularity theory; on the other hand, it is of practical
importance since it makes possible to locate transition points
without the need to compute the derivatives of the ground state
fidelity per lattice site with respect to the control
parameter~\cite{statement1}.  In contrast, the von Neumann entropy,
a bipartite entanglement measure, fails to distinguish degenerate
symmetry-breaking ground states.  We illustrate the general scheme
in terms of the quantum Ising model in a transverse magnetic field
and the spin $1/2$ XYX model in an external magnetic field. Here, it
is worth emphasizing that, although the scheme is applicable to
quantum lattice models in any spatial dimensions, we restrict
ourselves to quantum systems on an infinite-size lattice in one
spatial dimension. This is achieved by exploiting the infinite
matrix product state (iMPS) algorithm initiated by
Vidal~\cite{vidal1}. The extension to quantum lattice systems in two
and higher spatial dimensions, which requires to use the infinite
projected entangled-pair state (iPEPS) algorithm~\cite{iPEPS},  is
deferred to another publication~\cite{libo}.

 {\it Infinite matrix product state algorithm and spontaneous symmetry breaking.}
For quantum many-body systems on an infinite-size lattice in one
spatial dimension, Vidal~\cite{vidal1} has developed a variational
algorithm to compute their ground state wave functions based on
their MPS representations, which is a variant of the MPS
algorithm~\cite{vidal2,vidal3} on a finite-size lattice in one
spatial dimension. Here, we briefly recall the key ingredients of
the algorithm. Assume that the Hamiltonian is translationally
invariant, and consists of the nearest-neighbor interactions: $H
=\sum _i h^{[i, i+1]}$, with $h^{[i, i+1]}$ being the
nearest-neighbor two-body Hamiltonian density.  Attached to each
site is a three-indices tensor $\Gamma^s_{A\;lr}$ or
$\Gamma^s_{B\;lr}$, and to each bond a diagonal (singular value)
matrix $\lambda_A$ or $\lambda_B$, depending on the evenness and
oddness of the $i$-th site and the $i$-th bond, respectively. Here,
$s$ is a physical index, $s=1,\cdots,d$, with $d$ being the
dimension of the local Hilbert space, and $l$ and $r$ denote the
bond indices, $l,r=1,\cdots, \chi$, with $\chi$ being the truncation
dimension. The imaginary time evolution amounts to computing $ |\Psi
(\tau)\rangle = \exp(- H \tau) |\Psi (0)\rangle / |\exp(- H \tau)
|\Psi (0)\rangle |$.  For large enough $\tau$  and a generic initial
state $\Psi (0)\rangle$, it yields a good approximation to the
ground state wave function, as long as there is a gap in the
spectrum of the system. Following the Suzuki-Trotter
decomposition~\cite{suzuki}, the imaginary time evolution operator
is reduced to a product of two-site evolution operators acting on
sites $i$ and $i+1$: $U(i,i+1) = \exp(-h^{[i,i+1)]} \delta \tau )$,
$\delta \tau <<1$. Notice that, a two-site gate $U(i,i+1)$ renders
the state not in the form of a MPS and breaks the translational
invariance. The former is remedied by performing a singular value
decomposition of a matrix contracted from one $\Gamma^s_{A\;lr}$,
one $\Gamma^s_{B\;lr}$, one $\lambda_A$ and two $\lambda_B$'s, and
only the $\chi$ largest singular values are retained. This yields
the new tensors $\Gamma^s_{A\;lr}$,  $\Gamma^s_{B\;lr}$ and
$\lambda_A$, which are used to update the tensors for all the sites,
thus restoring the translational invariance under two site shifts.
Repeating this procedure until the ground state energy converges,
one may generate the systems's ground state wave functions in the
MPS representations.

Remarkably, for a system with symmetry-breaking orders, the iMPS
algorithm automatically produces degenerate ground states arising
from an SSB in the symmetry-broken phase, each of which breaks the
symmetry of the system. Moreover, the symmetry breakdown results
from the fact that an initial state has been chosen randomly. It is
worth mentioning that, for quantum lattice systems in one spatial
dimension, continuous symmetries cannot be spontaneously
broken~\cite{mw}, due to strong quantum
fluctuations~\cite{statement3}. Therefore, we shall restrict
ourselves to the discussion of quantum lattice systems with a
discrete symmetry group $Z_2$~\cite{statement2}.

{\it Bifurcations in the ground state fidelity per lattice site.}
Now consider a quantum many-body system, with a discrete symmetry
group $Z_2$, on an infinite-size lattice in one spatial dimension.
Assume that the system undergoes a continuous QPT with $Z_2$
symmetry spontaneously broken, when a control parameter $\lambda$
varies. According to the definition~\cite{zhou,zov}, the ground
state fidelity per lattice site, $d(\lambda_1, \lambda_2)$, is the
scaling parameter, which characterizes how fast the fidelity
$F(\lambda_1, \lambda_2) \equiv |\langle \Psi (\lambda_2) |\Psi
(\lambda_1) \rangle | $ between two ground states $|\Psi (\lambda_1)
\rangle$  and $|\Psi (\lambda_2) \rangle$ goes to zero when the
thermodynamic limit is approached. In fact, the ground state
fidelity $F(\lambda_1, \lambda_2)$ asymptotically scales as
$F(\lambda_1, \lambda_2) \sim  d(\lambda_1, \lambda_2)^L$, with $L$
the number of sites in a finite-size lattice. Remarkably, the ground
state fidelity per lattice site is well defined in the thermodynamic
limit, and satisfies the properties inherited from the fidelity
$F(\lambda_1, \lambda_2)$: (i) normalization $d(\lambda, \lambda)
=1$; (ii) symmetry $d(\lambda_1, \lambda_2) = d(\lambda_2,
\lambda_1)$; and (iii) range $0 \le d(\lambda_1, \lambda_2) \le 1$.

In the $Z_2$ symmetric phase, the ground state is non-degenerate,
whereas in the $Z_2$ symmetry-broken phase, two degenerate ground
states arise. Now let us see what this implies for the ground state
fidelity per lattice site, $d(\lambda_1, \lambda_2)$. If we choose
$\Psi (\lambda_2)$ as a reference state, with $\lambda_2$ in the
$Z_2$ symmetric phase, then  the ground state fidelity per lattice
site, $d(\lambda_1, \lambda_2)$, cannot distinguish two degenerate
ground states $|\Psi_\pm (\lambda_1) \rangle$ in the $Z_2$
symmetry-broken phase. Here,  $|\Psi_+ (\lambda_1) \rangle =P
|\Psi_- (\lambda_1) \rangle$, with $P$ being the operation
generating the symmetry group $Z_2$. This follows from the fact that
$\langle \Psi (\lambda_2) |\Psi_+ (\lambda_1) \rangle = \langle \Psi
(\lambda_2) |P|\Psi_+ (\lambda_1) \rangle = \langle \Psi (\lambda_2)
|\Psi_- (\lambda_1) \rangle$, for any large but finite size $L$.
However, if we choose $\Psi (\lambda_2)$ as a reference state, with
$\lambda_2$ in the $Z_2$ symmetry-broken phase, then $d(\lambda_1,
\lambda_2)$ is able to distinguish two degenerate ground states.
Therefore, for a given truncation dimension $\chi$, a pseudo phase
transition point $\lambda_\chi$ manifests itself as a
\textit{bifurcation point}~\cite{bifurcation}. An extrapolation to
$\chi = \infty$ determines the critical point $\lambda_c$.
Therefore, the pinch point, first introduced in
Refs.~\cite{zhou,zov} as an intersection of two singular lines to
characterize phase transition points, is identified as a bifurcation
point. The significance of this result lies in the fact that, on the
one hand, this establishes the connection between the fidelity
approach to QPTs and the singularity theory; on the other hand, it
is of practical importance since it makes possible to locate phase
transition points without the need to compute the derivatives of the
ground state fidelity per lattice site with respect to the control
parameter.

In contrast, the von Neumann entropy, a bipartite entanglement
measure, fails to distinguish degenerate symmetry-breaking ground
states.  This is due to the fact that the von Neumann entropy is
fully determined by the singular value matrices $\lambda_A$ and
$\lambda_B$, whereas all the information concerning an SSB is
encoded in the tensors $\Gamma^s_{A\;lr}$ and $\Gamma^s_{B\;lr}$.

{\it Models.} As an illustration, let us consider two quantum
systems with the symmetry group $Z_2$. The first is the quantum
Ising model in a transverse magnetic field on an infinite-size
lattice in one spatial dimension. It is described by the
Hamiltonian:
\begin{equation}
\small
  H=-\sum_{i=-\infty}^\infty \left(S^{[i]}_{x}S^{[i+1]}_{x}+ \lambda
  S^{[i]}_{z}\right), \label{ham1}
\end{equation}
where $S^{[i]}_{\alpha}$ ($\alpha=x,z$) are the Pauli spin operators
of the $i$-th spin $1/2$, and $\lambda$ is the transverse magnetic
field. The model is invariant with respect to the operation:
$S^{[i]}_{x}\rightarrow -S^{[i]}_{x}$ for all the sites
simultaneously, thus it enjoys the $Z_2$ symmetry. As is well known,
it undergoes a QPT, with a critical point at $\lambda_c
=1$~\cite{ising}.

The second is the spin $1/2$ XYX model in an external magnetic
field, with the Hamiltonian:
\begin{equation}
\small
  H=\sum_{i=-\infty}^\infty \left(S^{[i]}_{x}S^{[i+1]}_{x}+
  {\Delta}_{y}
  S^{[i]}_{y}S^{[i+1]}_{y}+ S^{[i]}_{z}S^{[i+1]}_{z}+h S^{[i]}_{z}\right), \label{ham2}
\end{equation}
where $S^{[i]}_{\alpha}$ ($\alpha=x,y,z$) are the Pauli spin
operators of the $i$-th spin $1/2$,  $\Delta_{y}$ denotes the
anisotropy in the internal spin space, and $h$ is an external
magnetic field. The model possesses a $Z_2$ symmetry, generated by
the operation: $S^{[i]}_{x}\rightarrow -S^{[i]}_{x}$ and
$S^{[i]}_{y}\rightarrow -S^{[i]}_{y}$.  Note that $\Delta_{y}< 1$
and $\Delta_{y}> 1$ correspond to easy-plane and easy-axis
behaviors, respectively. The ordered phase in the easy-plane
(easy-axis) case arises from an SSB along the $x (y)$ direction,
with a non-zero order parameter, i.e., the magnetization $\langle
S^{[i]}_{x}\rangle$  ($\langle S^{[i]}_{y}\rangle$) below the
critical field $h_c$. Here we shall choose $\Delta_{y}=0.25$, for
which it is critical at  $h = h_c $, with $h_c \sim 3.210(6)$ from
the quantum Monte Carlo simulation~\cite{xyx}.

\begin{figure}
 \includegraphics[width=0.48\textwidth]{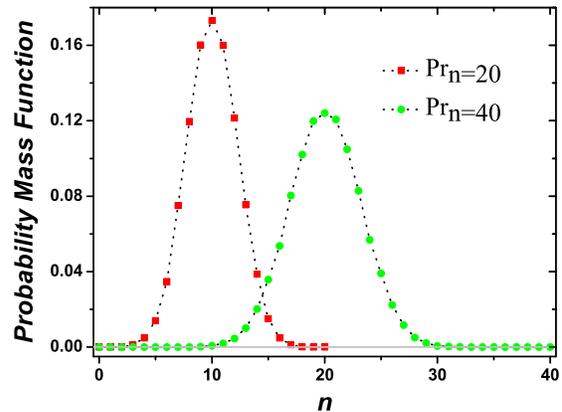}
\caption{(color online) The  probability mass function for the
quantum Ising model in a transverse magnetic field in the $Z_2$
symmetry-broken phase ($\lambda =1/2$). If a random variable $K$
follows the binomial distribution with parameters $n$ and $p$, then
the probability of getting exactly $k$ successes in $n$ trials is
given by the probability mass function: ${\rm Pr} \; (K=k) = C^k_n
p^k (1-p)^{n-k}$, for $k=0, 1, 2, \cdots, n$, where $C^k_n =n! /(k!
(n-k)!)$ is the binomial coefficient. Here, by a success we mean
that the order parameter $\langle S^{[i]}_{x} \rangle$ is positive.
Our data are presented for both $n=20$ and $n=40$, with $\chi =8$.
This confirms that the probability for getting the ground state with
the positive order parameter $\langle S^{[i]}_{x}\rangle$ each
simulation run is $p=1/2$. The same pattern occurs for other choices
of the truncation dimension $\chi$. Therefore, an SSB occurs in
classical simulations of quantum systems on an infinite-size lattice
in the context of the iMPS algorithm.} \label{FIG1}
\end{figure}

{\it Simulation results.} In Fig.~\ref{FIG1}, we present the
probability mass function for the quantum Ising model in a
transverse magnetic field in the $Z_2$ symmetry-broken phase
($\lambda =1/2$). Suppose a random variable $K$ follows the binomial
distribution with parameters $n$ and $p$, then the probability of
getting exactly $k$ successes in $n$ trials is given by the
probability mass function: ${\rm Pr} \; (K=k) = C^k_n p^k
(1-p)^{n-k}$, for $k=0, 1, 2, \cdots, n$, where $C^k_n =n! /(k!
(n-k)!)$ is the binomial coefficient. Here, by a success we mean
that the order parameter $\langle S^{[i]}_{x} \rangle$ is positive.
Our data are presented for both $n=20$ and $n=40$, with the
truncation dimension $\chi$ to be 8. This confirms that the
probability for getting the ground state with the positive order
parameter $\langle S^{[i]}_{x}\rangle$ each simulation run is
$p=1/2$. The same pattern occurs for other choices of the truncation
dimension $\chi$. Actually this is true for any model with $Z_2$
symmetry spontaneously broken. Therefore, our results demonstrate
that an SSB occurs in classical simulations of quantum systems on an
infinite-size lattice in the context of the iMPS algorithm. In
contrast, algorithms that simulate finite-size lattice systems are
forbidden to produce degenerate symmetry-breaking ground states,
since an SSB \textit{only} occurs in the infinite-size
(thermodynamic) limit.

\begin{figure}
\includegraphics[width=0.48\textwidth]{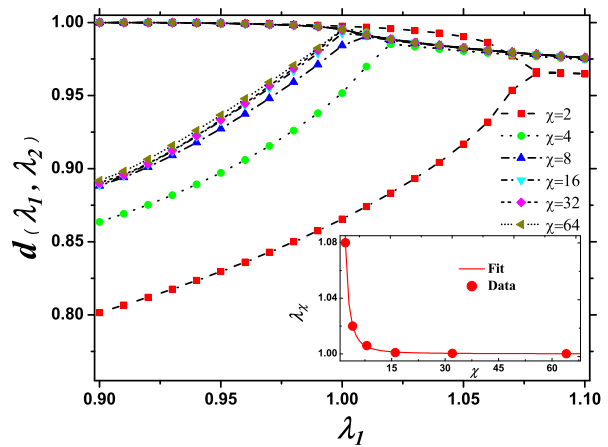}
 \caption{(color online) Main: The ground state fidelity per lattice site,
 $d(\lambda_1, \lambda_2)$, for the quantum Ising model in a transverse field.
 Here, the transverse magnetic field strength $\lambda$ is the control parameter.
If we choose $\Psi (\lambda_2)$ as a reference state, with
$\lambda_2$ in the $Z_2$ symmetry-broken phase (as shown here,
$\lambda_2 =0.9$), then  $d(\lambda_1, \lambda_2)$ is able to
distinguish two degenerate ground states, with a pseudo phase
transition point $\lambda_\chi$ as a \textit{bifurcation point}.
Inset: The critical point is determined from an extrapolation of the
pseudo phase transition point $\lambda_\chi$ for the truncation
dimension $\chi$. The fitting function is $\lambda_\chi=\lambda_c+a
\chi^{-b}$,  where $\lambda_c=1.00015$, with $a=0.31612$ and
$b=1.98565$. This indicates that we are able to locate the
transition point accurately, with moderate computational cost. The
accuracy may be further improved if the truncation dimension $\chi$
is increased.}
  \label{FIG2}
   \end{figure}

In Fig.~\ref{FIG2} we plot the ground state fidelity per lattice
site, $d(\lambda_1, \lambda_2)$, for the quantum Ising model in a
transverse field. Here, the transverse magnetic field strength
$\lambda$ is the control parameter.  If we choose $\Psi (\lambda_2)$
as a reference state, with $\lambda_2$ in the $Z_2$ symmetry-broken
phase (as shown here, $\lambda_2 =0.9$), then $d(\lambda_1,
\lambda_2)$ is able to distinguish two degenerate ground states,
with a pseudo phase transition point $\lambda_\chi$ as a
\textit{bifurcation point}~\cite{exact}. The critical value
$\lambda_c=1.00015$ is determined from an extrapolation of the
pseudo phase transition point $\lambda_\chi$ for the truncation
dimension $\chi$ (see the inset in Fig.~\ref{FIG2}), which is quite
close to the exact value $1$. Therefore, the iMPS algorithm enables
us to locate the transition point accurately from the computation of
$d(\lambda_1, \lambda_2)$, with moderate computational cost. We
stress that such a scaling for finite values of the truncation
dimension $\chi$ has been discussed for the von Neumann
entropy~\cite{tag}.

\begin{figure}
 \includegraphics[width=0.48\textwidth]{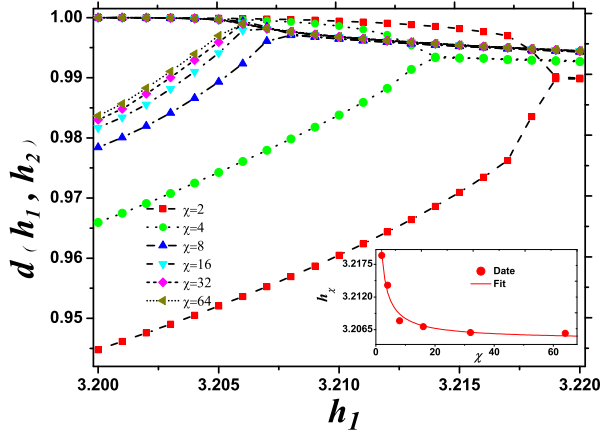}
\caption{(color online)  Main: The ground state fidelity per lattice
site, $d(h_1, h_2)$, for the spin $1/2$ XYX model in an external
magnetic field. Here, the external magnetic field $h$ is the control
parameter.  If we choose $\Psi (h_2)$ as a reference state, with
$h_2$ in the $Z_2$ symmetry-broken phase (as shown here, $h_2
=3.2$), then $d(h_1, h_2)$ is able to distinguish two degenerate
ground states, with a pseudo phase transition point $h_\chi$ as a
\textit{bifurcation point}. Inset: The critical point $h_c$ is
determined from an extrapolation of the pseudo phase transition
point $h_\chi$ for the truncation dimension $\chi$.  The fitting
function is $h_\chi =h_c + a \chi^{-b}$,  where $h_c =3.20471$, with
$a=0.02653$ and  $b=0.86003$.} \label{FIG3}
\end{figure}

We have also presented $d(h_1, h_2)$ for the spin $1/2$ XYX model in
an external magnetic field on an infinite-size lattice in
Fig.~\ref{FIG3}. Here, the external magnetic field $h$ is the
control parameter. If we choose $\Psi (h_2)$ as a reference state,
with $h_2$ in the $Z_2$ symmetry-broken phase (as shown here, $h_2
=3.2$), then $d(h_1, h_2)$ is able to distinguish two degenerate
ground states, with a pseudo phase transition point $h_\chi$ as a
\textit{bifurcation point}.  The critical point $h_c =3.20471$ is
determined from an extrapolation of the pseudo phase transition
point $h_\chi$ for the truncation dimension $\chi$, as seen from the
inset in Fig.~\ref{FIG3}.

 {\it Summary.} We have demonstrated that the iMPS algorithm may produce degenerate,
symmetry-breaking, ground states arising from an SSB, each of which
results from a randomly chosen initial state. It is shown that an
SSB is reflected as a bifurcation in  the ground state fidelity per
lattice site for quantum lattice systems undergoing QPTs with
symmetry-breaking orders. Conceptually, this establishes the
connection between the fidelity approach to QPTs and the singularity
theory. Practically, it is also important since it makes possible to
locate transition points without the need to compute the derivatives
of the ground state fidelity per lattice site with respect to the
control parameter, which is usually a formidable task. We
illustrated the general scheme in terms of the quantum Ising model
in a transverse magnetic field and the spin $1/2$ XYX model in an
external magnetic field on an infinite-size lattice in one spatial
dimension.

Finally, we point out that we may extend our investigation to
quantum lattice systems in two and higher spatial dimensions, which
requires the iPEPS algorithm~\cite{iPEPS}. This is currently under
active investigation~\cite{libo}.

{\it Acknowledgements.} The support from the National Natural
Science Foundation of China (Grant Nos: 10774197 and 10874252) and
the Natural Science Foundation of Chongqing (Grant No: CSTC,
2008BC2023) is acknowledged.



\begin{thebibliography}{10}

\bibitem{sachdev} S. Sachdev, Quantum Phase Transitions, Cambridge University
Press, 1999, Cambridge.

\bibitem{wen} X.-G. Wen, Quantum Field Theory of Many-Body Systems, Oxford
University Press, 2004, Oxford.

\bibitem{anderson} P.W. Anderson, Basic Notions of Condensed Matter
Physics,  Addison-Wesley: The Advanced Book Program, 1997, Reading,
Mass.

\bibitem{coleman} S. Coleman,  An Introduction to Spontaneous Symmetry
Breakdown and Gauge Fields, Laws of Hadronic Matter,  Ed. A.
Zichichi, Academic, 1975, New York.

\bibitem{book} See, e.g., L. Amico, R. Fazio, A. Osterloh, and V. Vedral, Rev.
Mod. Phys. \textbf{80}, 517 (2008) and references therein.

 \bibitem{zp} P. Zanardi and N.
Paunkovi\'{c}, Phys. Rev. E  \textbf{74}, 031123 (2006).

 \bibitem{zhou} H.-Q. Zhou and J.P. Barjaktarevi$\check{\rm c}$,
J. Phys. A: Math. Theor. \textbf{41} 412001 (2008); H.-Q. Zhou,
J.-H. Zhao, and B. Li, J. Phys. A: Math. Theor. \textbf{41} 492002
(2008); H.-Q. Zhou, arXiv:0704.2945.

\bibitem{zov} H.-Q. Zhou, R. Or\'{u}s, and G. Vidal,
Phys. Rev. Lett. \textbf{100}, 080602 (2008).

\bibitem{fidelity} P. Zanardi, M. Cozzini, and P. Giorda, J. Stat. Mech. L02002, (2007);
N. Oelkers and J. Links, Phys. Rev. B \textbf{75}, 115119 (2007); M.
Cozzini, R. Ionicioiu, and P. Zanardi, Phys. Rev. B \textbf{76},
104420 (2007); L. Campos Venuti and P. Zanardi, Phys. Rev. Lett.
\textbf{99}, 095701 (2007);
 W.-L. You, Y.-W. Li, and S.-J. Gu, Phys. Rev. E \textbf{76},
022101 (2007); S. J. Gu \textit{et al.}, Phys. Rev. B \textbf{77},
245109 (2008); M. F. Yang, Phys. Rev. B \textbf{76}, 180403(R)
(2007); Y. C. Tzeng and M.F. Yang, Phys. Rev. A 77, 012311 (2008);
J. O. Fj{\ae}restad, J. Stat. Mech. P07011 (2008).



\bibitem{nielsen} M.A. Nielsen and I.L. Chuang, Quantum Computation and Quantum
Information (Cambridge University Press, 2000, Cambrige).

\bibitem{statement} This is even valid for thermal phase transitions if we extend the fidelity
notion from pure states to mixed states to accomodate thermal
fluctuations (see the first reference in~\cite{zhou}). In fact, the
logrithmic function of the fidelity per lattice site for two thermal
mixed states corresponding to different temperatures reduces to
nothing but the free energy, if other non-thermal control parameters
are kept fixed. This implies that the singularities in the fidelity
per lattice site coincide with those in the free energy, thus
showing the equivalence between the fidelity approach and the
conventional one to thermal phase transitions.

\bibitem{zhou1} J.-H. Zhao and H.-Q. Zhou, arXiv:0803.0814.

\bibitem{zhou2} H.-L. Wang, J.-H. Zhao, B. Li, and H.-Q. Zhou, arXiv:0902.1670.


\bibitem{zhou3} H.-Q. Zhou, arXiv:0803.0585.

\bibitem{bifurcation} Bifurcation theory studies and classifies phenomena characterized by a sudden change in
behaviors arising from a small variation in a control parameter. For
a review, see, J.D. Crawford, Rev. Mod. Phys. \textbf{63}, 991
(1991). See also J. Araki \textit{et al.}, Proc. R. Soc. Lond.
\textbf{A345}, 413 (1975) for a discussion about the spontaneously
broken symmetry and the cusp catastrophe.


\bibitem{statement1} Numerically, it is a formidable task to compute
the derivatives of the ground state fidelity per lattice site with
respect to the control parameter, due to stringent accurary
requirements.

\bibitem{vidal1} G. Vidal, Phys. Rev. Lett. 98, 070201 (2007).


\bibitem{iPEPS} J. Jordan \textit{et al.}, Phys. Rev. Lett. {\bf 101}, 250602 (2008).


\bibitem{libo} B. Li \textit {et al.}, in preparation.


\bibitem{vidal2} G. Vidal, Phys. Rev. Lett. 93, 040502 (2004).

\bibitem{vidal3} G. Vidal, Phys. Rev. Lett. 91, 147902 (2003).


\bibitem{suzuki} M. Suzuki, Phys. Lett. A146, 319 (1990).





\bibitem{mw} N. D. Mermin and H. Wegner, Phys. Rev. Lett.  \textbf{17}, 1133 (1966).









\bibitem{statement3} However, the finiteness of the truncation dimension $\chi$ in the iMPS
algorithm automatically leads to infinite degenerate ground states,
each of which breaks the continuous symmetry. This results in the
so-called \textit{pseudo} continuous SSB~\cite{zhou2}, for quantum
lattice systems with a continuous symmetry group, e.g., $U(1)$. In
addition, the extent to which the symmetry is spontaneously broken
may be quantified by introducing a pseudo-order parameter that must
be scaled down to zero, in order to be consistent with the
Mermin-Wegner theorem.



\bibitem{statement2} The extension of our discussion to quantum lattice systems with other discrete groups, e.g., $Z_N$,  is
straightforward.

\bibitem{ising} E. Lieb, T. Schultz, and D. Mattis, Ann. Phys. 60, 407 (1961); P. Pfeuty, Ann. Phys. 57, 79 (1970).

\bibitem{xyx} J. Kurmann \textit {et al.}, Physica \textbf{A112}, 235 (1982); D. V. Dmitriev \textit {et al.},
J. Exp. Th. Phys. \textbf{95}, 538 (2002); T. Roscilde, \textit {et
al.}, Phys. Rev. Lett. \textbf{93}, 167203 (2004).

\bibitem{exact} We emphaize that it is difficult, if not impossible,
to figure out the bifurcation in the ground state fidelity per
lattice site for the quantum Ising model in a transverse field from
the exact solution of the model~\cite{ising}. In fact,  the
Jordan-Wigner transformation, used to solve the model, changes the
boundary conditions.  Therefore, it affects the ground state
fidelity per lattice site  for finite-size systems, but not in the
infinite-size (thermodynamic) limit.




\bibitem{tag} L. Tagliacozzo, Thiago. R. de Oliveira, S. Iblisdir, and J. I.
Latorre,  Phys. Rev. \textbf{B78}, 024410 (2008).







 \end{thebibliography}
\end{document}